\documentclass[aps,preprint,floatfix,amssymb,amsmath,prb,endfloats]{revtex4}
\usepackage{graphicx}
\usepackage{bm}
\begin{document}
\title{ 
Inherent flexibility determines the
transition mechanisms of the EF-hands of Calmodulin
}

\author{Swarnendu Tripathi and John J. Portman}

\affiliation{
Department of Physics, Kent State University, Kent, OH 44242
}

\begin{widetext}
\begin{abstract} 
We explore how inherent flexibility of a protein molecule influences
the mechanism controlling the kinetics of allosteric transitions using
a variational model inspired from work in protein folding.  The
striking differences in the predicted transition mechanism for the
opening of the two domains of calmodulin (CaM) emphasizes that
inherent flexibility is key to understanding the complex
conformational changes that occur in proteins.  In particular, the
C-terminal domain of CaM (cCaM) which is inherently less flexible than
its N-terminal domain (nCaM) reveals ``cracking" or local partial
unfolding during the open/closed transition. This result is in harmony
with the picture that cracking relieves local stresses due to
conformational deformations of a sufficiently rigid protein.  We also
compare the conformational transition in a recently studied ``even-odd"
paired fragment of CaM. Our results rationalize the different relative
binding affinities of the EF-hands in the engineered fragment compared
to the intact ``odd-even" paired EF-hands (nCaM and cCaM) in terms of
changes
in flexibility along the transition route.  Aside from
elucidating general theoretical ideas about the cracking mechanism,
these studies also emphasize how the remarkable intrinsic plasticity
of CaM underlies conformational dynamics essential for its diverse
functions.   
\end{abstract}  
\end{widetext}                                                               

\date{\today} 

\maketitle



\section*{INTRODUCTION}

To understand protein function, it is often essential to characterize
large conformational changes that occur upon ligand binding.  Within
the population-shift mechanism of allosteric conformational
change,\cite{weber:72} the bound complex is formed
when a ligand selects and stabilizes a weakly populated conformational
ensemble from the kinetically accessible states of the unbound folded
protein. Consequently, the kinetics of large scale
conformational transitions between two meta-stable states
are determined largely by the
inherent conformational dynamics within the folded state free energy
basin.  Such conformational dynamics imply an inherent flexibility or
``intrinsic plasticity'' of the folded state. In this paper, we focus
on how this inherent flexibility of a protein molecule influences the
mechanism controlling the kinetics of allosteric transitions. There are a couple
of possible scenarios for the transition mechanism in terms of changes in conformational
flexibility. One possibility is that
the flexibility adjusts smoothly to the conformation
deformation between two specific meta-stable states in the free energy
surface. In this case, the inherent flexibility of the protein 
remains relatively constant if the meta-stable states have similar
flexibilities or else changes smoothly between the flexibilities of the
meta-stable states if the flexibilities are different.
Another possible mechanism, called
``cracking'',\cite{miyashita:wolynes:03} has recently been proposed as an
alternative mechanism that may be important in some conformational
transitions.  In terms of conformational flexibility, cracking
involves non-monotonic changes in flexibility along the transitions
route.  In particular, the flexibility of specific regions of the
protein may transiently increase through local unfolding. As
explained in Ref.\onlinecite{miyashita:wolynes:03}, local unfolding can relieve
specific areas of high stress during conformational deformations that
would result in high free energy barriers if the protein remained
uniformly folded throughout the transition.

In the model developed to explore this idea, cracking is introduced
directly into the formalism as means to incorporate nonlinear
elasticity in an otherwise harmonic description of conformational
fluctuations (i.e., normal
modes).~\cite{miyashita:wolynes:03,miyashita:wolynes:05} This work clearly shows
that local unfolding can dramatically lower predicted free energy
barriers between local free energy minima and hence facilitates faster
kinetics. In contrast, the variational model developed in the present
work does not assume cracking from the outset. The evidence of
cracking here arises entirely from the analysis of the local changes
in flexibility along transition routes predicted by an inherently
non-linear model of conformational transitions. Our results are thus
an independent verification of the ideas and formalism developed in the model
presented in Ref.\onlinecite{miyashita:wolynes:03,miyashita:wolynes:05}.

Cracking does not occur in all conformational transitions, but even
when it does, the average conformations of the local minima in the
free energy may show little signs that local unfolding is
involved.\cite{miyashita:wolynes:03}  Instead, cracking is a subtle
consequence of the nature of the conformational
deformation required to connect the two states.  
Since cracking is akin to folding of the whole protein albeit
on a constrained and local scale, it is reasonable to ask if some
insight appropriated from the success of the energy landscape theory
of protein folding\cite{bryngelson:wolynes:95} can help to anticipate when cracking is
likely to be important in conformational transitions between two
distinct folded conformations.
One motivation for the work in this paper is to investigate if the
cracking mechanism of conformational transitions, like the mechanism
for folding of two state proteins,\cite{onuchic:wolynes:04}
is determined by
structural topology of the two meta-stable states. To address this
question we study the conformational transitions of the open and closed
conformations of the two homologous domains of calmodulin (CaM) as
well as a fragment of CaM involving parts of both domains.  Aside from
general theoretical implications, the results presented in this paper
are interesting from the point of view of understanding the well-known
and intensely studied thermodynamic differences between the two
domains of CaM.\cite{tsalkova:privalov:85,linse:forsen:91}

Our approach is based on a coarse-grained variational model previously
developed to characterize protein folding.~\cite{portman:wolynes:98}
This model is in harmony with several recent coarse-grained
simulations based on a folded-state biased interaction potential that
interpolates between the contact maps of the meta-stable
states.~\cite{zuckerman:04,best:hummer:05,maragakis:karplus:05,okazaki:wolynes:06,whitford:onuchic:07,chu:voth:07} 
Interestingly some of these recently proposed coarse-grained model for protein 
conformational transition can also capture the cracking or local partial unfolding 
during transition~\cite{best:hummer:05,okazaki:wolynes:06,whitford:onuchic:07}.
In this paper we compare the detailed predicted transition routes for
the open/closed conformational change of the C-terminal domain of
Calmodulin (cCaM) with that of the N-terminal domain of CaM (nCaM).
The open and closed state of each domain is shown in Fig.~\ref{fig:cam_domains}(a--b).
We find
that although the nCaM and cCaM have similar structures,
the calculated transition routes predict that the
cracking is essential only in cCaM. It is well
established that the two domains differ in flexibility as well as binding
affinity for Ca$^{2+}$.\cite{tsalkova:privalov:85,linse:forsen:91} These properties are not determined
by topology (since they are the same) but ultimately by more subtle
structural differences encoded in the sequence of the two domains. 
Our results suggest that nCaM is
flexible enough that cracking is not necessary to relieve stress
during the conformational change, while the cCaM is relatively rigid
so that cracking is involved in essentially the same conformational
change.  In addition, we investigate the conflicting Ca$^{2+}$ binding
mechanism in the ``odd-even" and ``even-odd" paired EF-hands between
the two domains of CaM [shown in Fig.~\ref{fig:cam_domains}(c)]. 
Our study indicates that 
the highly flexible central linker region locally unfolds before folding into a stable 
$\alpha$-helix. Our predicted transition mechanism rationalizes
the sequential Ca$^{2+}$ binding in CaM2/3.\cite{lakowski:mcintosh:07} in terms of the
differences in conformational flexibility of Ca$^{2+}$-binding
loops in CaM2/3. Finally, we address how our comparative study of
conformational transitions in ``odd-even" and ``even-odd" paired
EF-hands reveals the interplay between intrinsic plasticity, target
binding affinity, and function of CaM.

\section*{RESULTS}

\begin{figure}
   \begin{center}
      \includegraphics[width=4.5in]{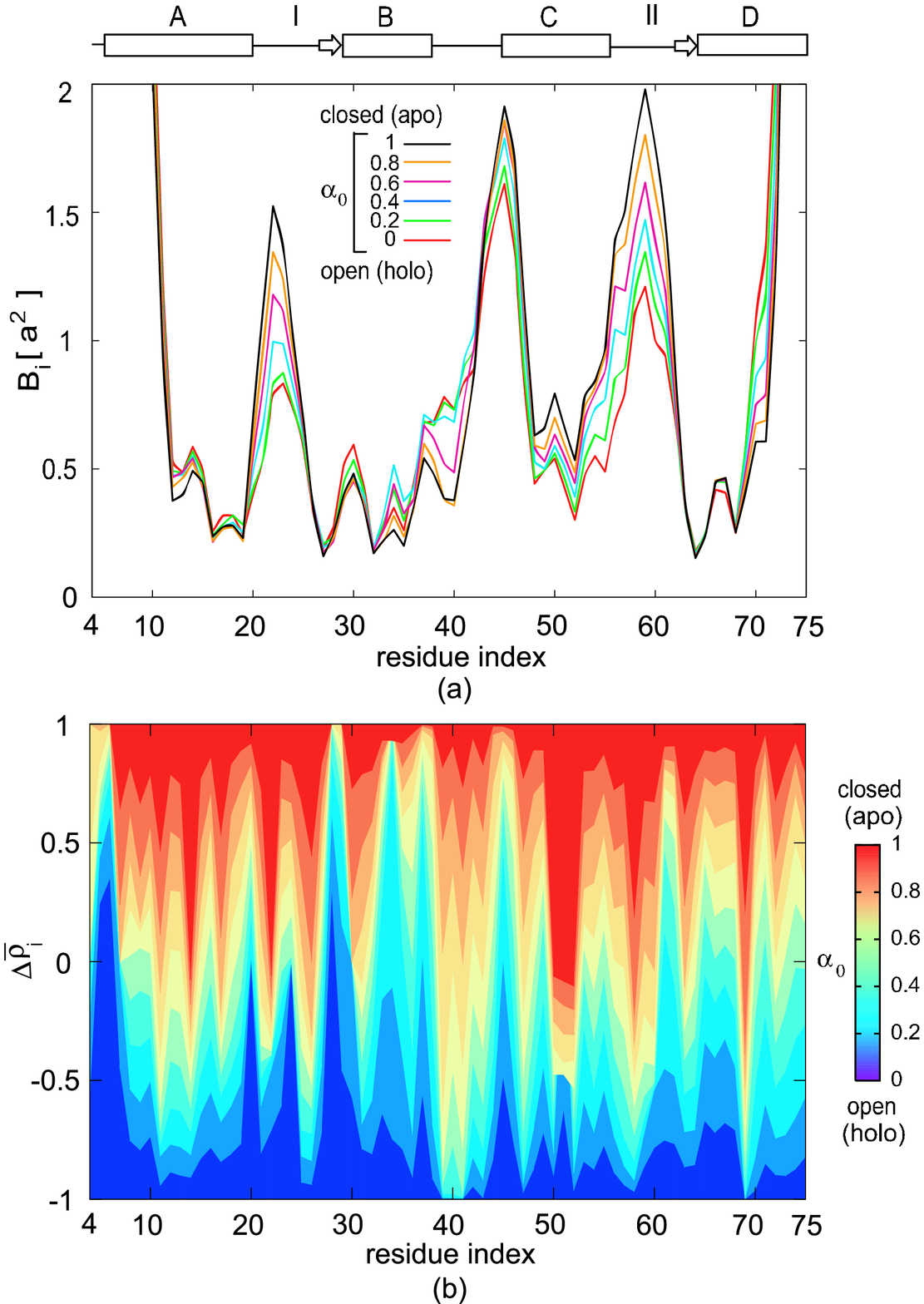}
      \caption{}
      \label{fig:B_rho}
   \end{center}
\end{figure}

\begin{figure}
   \begin{center}
      \includegraphics[width=4.5in]{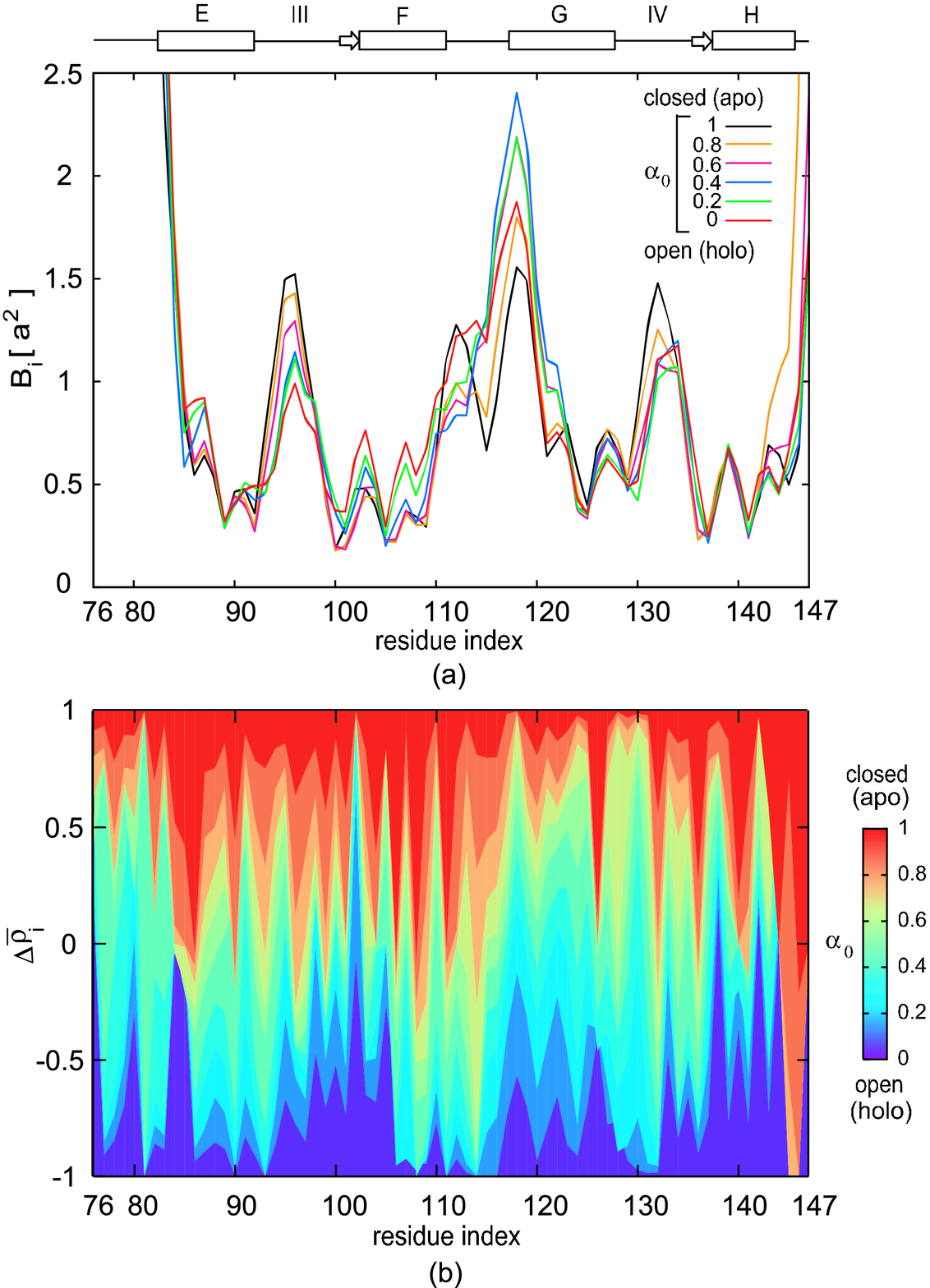}
      \caption{}
      \label{fig:B_rho_1}
   \end{center}
\end{figure}

\textbf{Conformational Flexibility of the CaM Domains.}~ We first
compare the inherent flexibility of the two domains of CaM by
calculating the mean-square fluctuations $\mathcal{B}_i$ for each
residue in the nCaM and cCaM for the open/closed conformational
transitions of each domain.  Fig.~\ref{fig:B_rho}(a) shows
$\mathcal{B}_i$ along the open/closed conformational transition for
each residue of nCaM from our very recent
studies.~\cite{tripathi:portman:08} The magnitude of the fluctuations
for each residue of cCaM for the open/closed transition route is shown
in Fig.~\ref{fig:B_rho_1}(a).  Comparison of Fig.~\ref{fig:B_rho}(a)
and ~\ref{fig:B_rho_1}(a) show that although the binding loops and
helix-linker in both domains are very flexible, the apo (closed)-nCaM
is inherently more flexible than apo (closed)-cCaM. For nCaM,
the flexibility of the the binding loops I and
II and the B/C helix-linker between EF-hands 1 and 2 decreases upon domain opening. Similarly, the
flexibility of the Ca$^{2+}$ binding loops III and IV decreases
considerably during the opening of cCaM. The fluctuations of the F/G helix-linker
between EF-hands 3 and 4 has different behavior along the transition route. 
The flexibility of this region of the protein
increases to a relatively high flexibility in the
intermediate states between the open and closed conformations
before reducing the flexibility of the folded open conformation.
(Fig.~\ref{fig:B_rho_1}(a)). 
We also note that
the F/G helix-linker in the open structure
is more flexible than in the closed structure. 
The calculated conformational
flexibility suggests that unlike the B/C helix-linker in the apo
(closed)-nCaM the F/G helix-linker in the apo (closed)-cCaM is
relatively less flexible. The increase and then decrease in flexibility of
the F/G helix-linker during the domain opening of cCaM indicates there is
cracking or local partial unfolding in this less flexible F/G
helix-linker which facilitates the conformational change by increasing
its flexibility near the transition state ensemble. [See the change in
fluctuations of the F/G helix-linker during domain opening in
Fig. ~\ref{fig:B_rho_1}(a)] The model also predicts that binding
loop II in CaM has the highest flexibility in comparison with other
binding loops. [Compare the flexibility of the binding loops in
Fig.~\ref{fig:B_rho}(a) and ~\ref{fig:B_rho_1}(a).] This result agrees
with a recent molecular dynamics simulation study of CaM D129N mutant,
where the binding loop II consistently exhibited higher mobility than
the other three Ca$^{2+}$ binding loops.~\cite{likic:gooley:03}

\begin{figure}
   \begin{center}
      \includegraphics[width=6in]{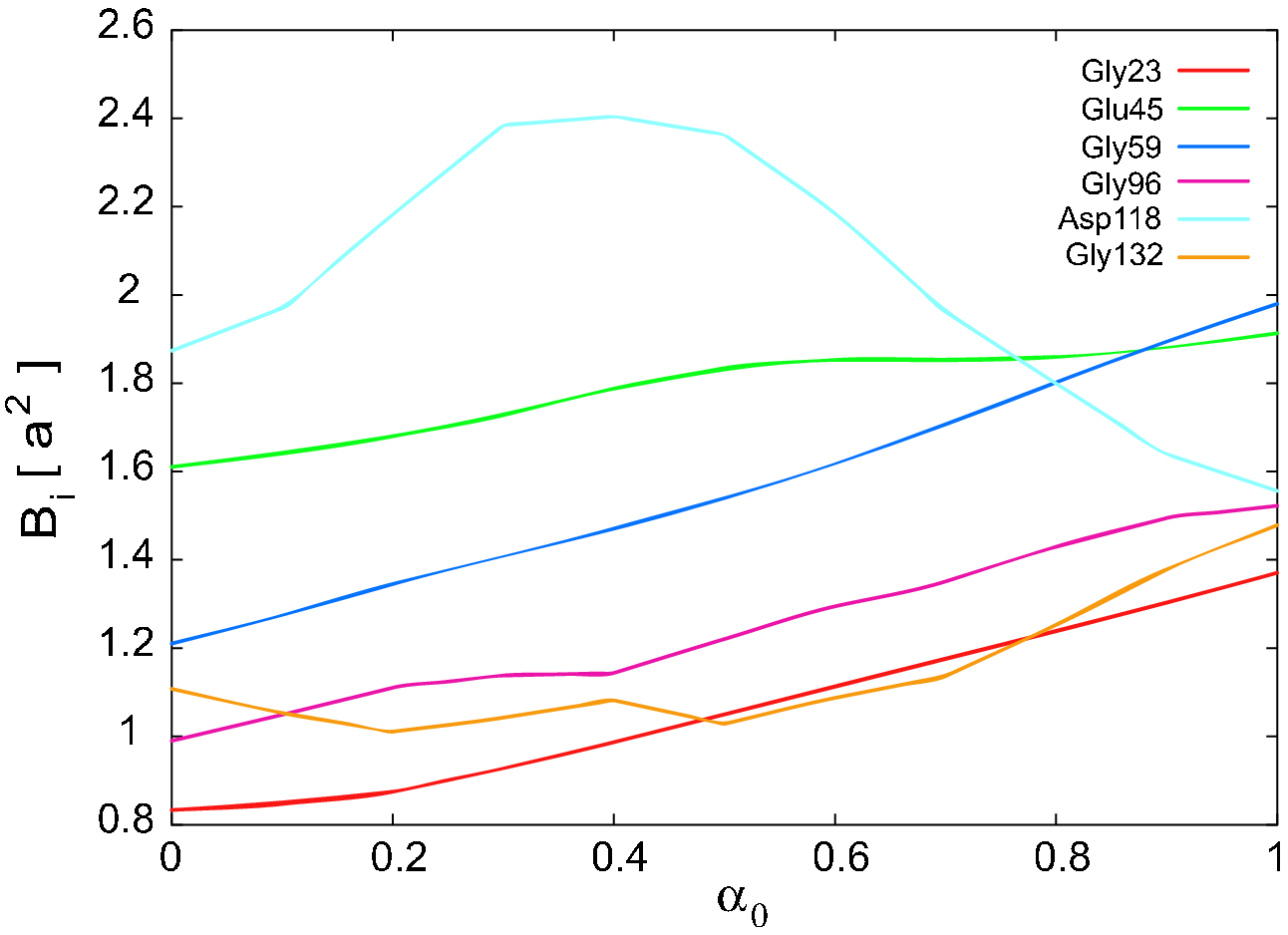}
      \caption{}
      \label{fig:B_domains}
   \end{center}
\end{figure}

\begin{figure}
   \begin{center}
      \includegraphics[width=6in]{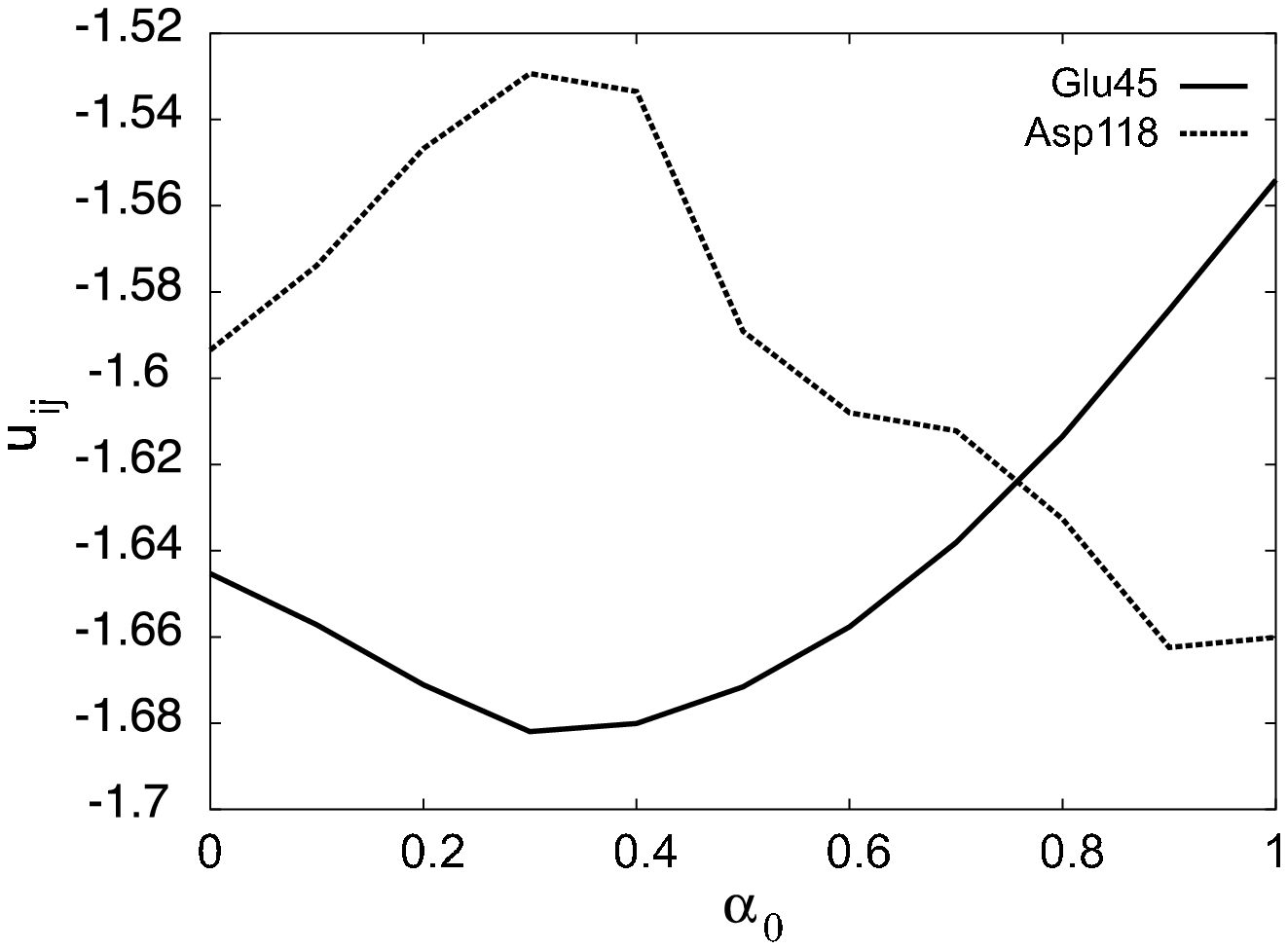}
      \caption{}
      \label{fig:uij_domains}
   \end{center}
\end{figure}

\textbf{Cracking in the Conformational Transition of cCaM.}~ To compare
the conformational flexibility of the two CaM domains in more detail
we have plotted the change in fluctuations $\mathcal{B}_i$ for some
specific residues in Fig.~\ref{fig:B_domains}. Each binding loop of CaM 
has 12 residues with a conserved Glutamate (Glu) residue in the
4$^{th}$ position of the loops.  In Fig.~\ref{fig:B_domains} we 
show Glu23 and Glu59 from binding loops I and II of nCaM,
respectively.  Also, Glu96 and Glu132 from binding loops III and IV of
cCaM, respectively. In Fig.~\ref{fig:B_domains}, we also included
the Glycine (Gly) and Aspartic (Asp) residues from the helix-linker at
position 45 and 118 of the nCaM and cCaM, respectively. [Note that all
these six residues show sharp peaks in Fig.~\ref{fig:B_rho}(a) and
~\ref{fig:B_rho_1}(a).] Residue Asp118 of the F/G helix-linker from
cCaM shows very different behavior relative to the rest of the
residues in Fig.~\ref{fig:B_domains}. Residue Asp118 shows highest
$\mathcal{B}_i$ near the transition state at $\alpha_0 = 0.4$ whereas
$\mathcal{B}_i$ of all the other residues shown decrease monotonically during
closed to open transition with the closed conformation having the largest
fluctuations. This result supports that this increase in
flexibility of the helix-linker region in cCaM near the transition
state is due to some local transient unfolding or
cracking. For further analysis of cracking in the cCaM we have also
compared the contact pair potential energy $u_{ij}$ for residues Glu45
and Asp118 from the helix-linker of nCaM and cCaM respectively along
the transition route. Fig.~\ref{fig:uij_domains} shows that for
residue Asp118, $u_{ij}$ increases in the transition state and
therefore cracking occurs. On the other hand, $u_{ij}$ for residue
Glu45 from the helix-linker of nCaM deceases in the transition
state. This striking difference in $u_{ij}$ implies that the increase
in contact energy of the F/G helix-linker during opening of cCaM leads
to some transient cracking in this region near the transition state
which further enhances the inherent flexibility of this linker during
the open/closed conformational change.

\textbf{Open/Closed Transition Mechanism of the CaM Domains.}~ To further
elucidate the predicted conformational transition mechanisms of domain opening in CaM,
we consider a structural order parameter that measures the similarity
to the open (holo) or closed (apo) state conformations,
$\Delta\overline{\rho_i}$, described in the Methods section.
This order parameter is defined such that $\Delta\overline{\rho_i}=1$
corresponds to the closed conformation and
$\Delta\overline{\rho_i}=-1$ corresponds to the open conformation of
nCaM or cCaM. The transition routes illustrated by $\Delta\overline{\rho_i}$ for each residue
of nCaM and cCam are shown in Fig.~\ref{fig:B_rho}(b) and Fig.~\ref{fig:B_rho_1}(b),
respectively.
Fig.~\ref{fig:B_rho_1}(b) illustrates the conformational
transition of cCaM in terms of $\Delta\overline{\rho_i}$ for each
residue.  As shown in Fig.~\ref{fig:B_rho_1}(b), the model predicts that 
the structural change in binding loop II occurs earlier than binding loop III
during the domain opening of cCaM.
[See the sharp change in color near residue number 130 in
Fig.~\ref{fig:B_rho_1}(b)].  Also, helix G has an earlier
conformational transition than helix F (Fig.~\ref{fig:B_rho_1}(b)). This
sequence is quite different from nCaM as shown in 
Fig.~\ref{fig:B_rho}(b). For nCaM the structural change in
helix B is predicted to be earlier than the structural change in helix C for the closed to open
conformational transition. Finally, when we compared the structural
change in the helix-linker of the two domains, our results show F/G
helix-linker in the cCaM has abrupt transition near the open state.
[See the transition near residue number 115 in
Fig.~\ref{fig:B_rho_1}(b).] This may have some implication for
cracking in the F/G helix-linker region of cCaM. In contrast, the
transition in the B/C helix-linker in nCaM is more gradual (see
Fig.~\ref{fig:B_rho}(b)).

\textbf{Conformational Flexibility and Cracking of the CaM2/3
Fragment.}~ 
The engineered ``even-odd" EF-hands paired CaM2/3 fragment (composed
of EF-hands 2 and 3) has been shown recently to have distinct
transition characteristics from the ``odd-even" EF-hands paired nCaM
and cCaM.~\cite{lakowski:mcintosh:07} NMR spectroscopy
has shown that Ca$^{2+}$-free (apo) CaM2/3 does not have a stable
folded structure but rather shows characteristics of a molten globule
state. Ca$^{2+}$ binding induces the folding of this ``even-odd"
paired EF-hand motifs CaM2/3 and the Ca$^{2+}$-bound (holo) CaM2/3
adopts a similar structure as holo-nCaM or cCaM (see PDB ID {\em
2hf5}).~\cite{lakowski:mcintosh:07} In contrast to the Ca$^{2+}$-CaM
(holo-CaM) structure, the Ca$^{2+}$-CaM2/3 structure does not have a
stable helix in the central linker region. In our model, we study the
two meta-stable conformations, apo-CaM2/3 and holo-CaM2/3 were taken
directly from the folded apo-CaM (PDB ID {\em 1cfd}) and holo-CaM (PDB ID 
{\em 1cll}) respectively. Here, we focus mainly on the conformational
flexibility and transition of the central linker to a stable
$\alpha$-helix. This central linker of CaM was also previously studied
by MD simulation.~\cite{spoel:vogel:96} Even though our model does not
accommodate all the relevant conditions of the NMR experiment reported
in Ref.~\onlinecite{lakowski:mcintosh:07}, the model does capture
certain aspects of the the apo/holo conformational transition of this
CaM2/3 fragment. In particular, the model predicts a apo-CaM2/3 to
the holo-CaM2/3 conformational transition mechanism that agrees 
well with the sequential Ca$^{2+}$ binding mechanism of CaM2/3 suggested
by the NMR measurements.
  
\begin{figure}
   \begin{center}
      \includegraphics[width=4.5in]{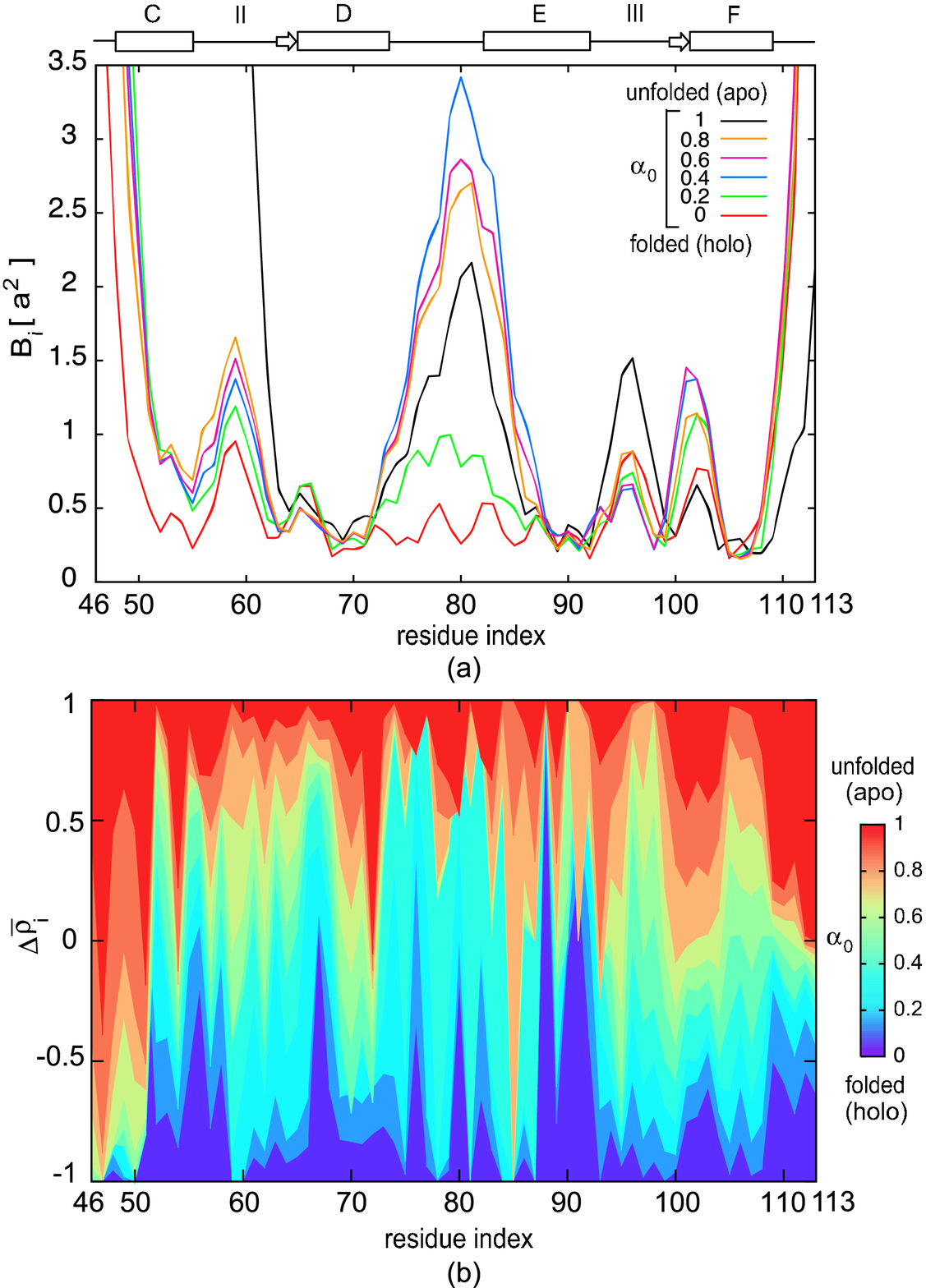}
      \caption{}
      \label{fig:B_rho_2}
   \end{center}
\end{figure}

The magnitude of the fluctuations of each residue in CaM2/3 [shown in
Fig.~\ref{fig:B_rho_2}(a)] suggests that the change in conformational
flexibility of binding loops II and III of CaM2/3 are very
different. In particular, helix C and binding loop II are highly
flexible in the apo conformation of CaM2/3. This very high intrinsic
flexibility of apo-CaM2/3 may account for the molten globule state
characteristics described in Ref.~\onlinecite{lakowski:mcintosh:07}.
Fig.~\ref{fig:B_rho_2}(a) also shows that binding loop II is more
flexible than loop III. This result is in harmony with the NMR
measurements of Ca$^{2+}$ binding in
CaM2/3~\cite{lakowski:mcintosh:07} that shows Ca$^{2+}$ binds to loop
III with higher affinity (lower Ca$^{2+}$ concentration) than loop II.
Our results also indicate that the flexibility of binding loop III
decreases monotonically along the transition route from the apo-CaM2/3
to the holo-CaM2/3 structure.  In contrast, the flexibility of helix F
increases along the transition route from apo to holo transition of
CaM2/3 fragment. In particular, N-terminal part (close to binding loop
III) of helix F partially unfolds when adopting the holo-CaM2/3
conformation.  We also notice in Fig.~\ref{fig:B_rho_2}(a) that the
domain linker between EF-hands 2 and 3 (residue number 74-81) during
the apo to holo transition has a relatively large change in
conformational flexibility. The fluctuation amplitude $\mathcal{B}_i$
of this linker increases in the intermediate states and then decrease
abruptly as the helix forms near the holo state.  The local unfolding
of the domain linker enhances the flexibility of this linker region
dramatically, first relaxing in structure and then stabilized in
the holo-CaM2/3 structure as a rigid $\alpha$-helix conformation.
This local unfolding signaled by increase and decrease in flexibility
is similar to the cracking exhibited in cCaM.

Local intermediate unfolding (i.e., cracking ) of the domain linker
during the apo to holo transition of CaM2/3 can also be seen through
change in the pair potential energy $u_{ij}$ for the Aspartic (Asp)
residue from the linker at position 80 along the transition route as
shown in Fig.~\ref{fig:uij_cam2/3}.  During the apo to holo transition
of CaM2/3 the energy of residue Asp80 increases from the apo-CaM2/3
structure ($\alpha = 1$) and decreases abruptly at the transition
state ($\alpha_0 = 0.4$) to the minimum at holo-CaM2/3 structure
($\alpha_0 = 0$) as the helix forms . This increase and then sharp
decrease in the contact energy of residue Asp80 signals cracking
(local unfolding and refolding) of the linker in CaM2/3 to a stable
$\alpha$-helix. We note also that the energy relative to to the
holo-CaM2/3 energy of Asp80 is higher than that of Asp118 from the F/G
helix-linker of cCaM.  This emphasizes the importance of cracking in
the apo-CaM2/3 to holo-CaM2/3 transition.

\textbf{Conformational Transition Mechanism of CaM2/3 Fragment.}~For
CaM2/3, the order parameter $\Delta\overline{\rho_i}=1$ corresponds to
the apo state and $\Delta\overline{\rho_i}=-1$ corresponds to the holo
state of CaM2/3.  The abrupt and early change in
$\Delta\overline{\rho_i}=1$ for residues in the domain linker
(sequence number 74-81) illustrated in Fig.~\ref{fig:B_rho_2}(b)
clearly shows the transition of the flexible inter domain linker to a
rigid $\alpha$-helix as discussed above in terms of the fluctuations
$\{\mathcal{B}_i\}$. Fig.~\ref{fig:B_rho_2}(b) also indicates that the
structural change in binding loop III is initiated earlier than the
structural change in binding loop II. Also, the conformational change
in binding loop II is much more gradual than that of binding loop III.
Similar to the identification of the Ca$^{2+}$ binding affinity with
the changes in inherent flexibility, the early structural change
in binding loop III may imply the stepwise binding of Ca$^{2+}$ in
CaM2/3 with loop III having a higher binding affinity than loop II.

\section*{DISCUSSION}

CaM, a small (148-residue) Ca$^{2+}$-binding protein with very high
plasticity, may be an ideal system to demonstrate flexibility influenced
conformational transitions. The
protein consists of structurally similar N- and C-terminal globular
domains connected by a flexible tether also known as central or interdomain 
linker.~\cite{barbato:bax:92} The Ca$^{2+}$ induced structural
rearrangement in CaM result in the solvent exposure of large
hydrophobic surface responsible for molecular recognition of various
cellular targets.~\cite{meador:quiocho:93}

While similar in structure and fold, the two CaM domains are quite
different in terms of their flexibility, melting temperatures, and
Ca$^{2+}$-binding
affinities.~\cite{tsalkova:privalov:85,linse:forsen:91} In the two
homologous (46\% sequence identity) domains of CaM, Ca$^{2+}$ binding
occurs sequentially. First, in the binding sites of cCaM and then in
the binding sites of nCaM.~\cite{linse:forsen:91} In spite of large
cooperativity in the Ca$^{2+}$ binding process within each domain the
two domains reflect different Ca$^{2+}$ and target affinities.  The
N-terminal pair of EF-hands binds to Ca$^{2+}$ ions with much lower
affinity than the C-terminal EF-hands pair.~\cite{linse:forsen:91}
Nuclear Magnetic Resonance (NMR)~\cite{chou:bax:01,evenas:akke:99} and
molecular-dynamics (MD) simulation~\cite{vigil:garcia:01} studies have
shown that the Ca$^{2+}$-bound nCaM is considerably less open than the
cCaM. This was not observed in the X-ray crystal structure of the
protein. Experimentally it has been shown that the more-conserved cCaM
has a greater affinity for Ca$^{2+}$ and some CaM targets, whereas the
nCaM is less specific in its choice of target
motif.~\cite{bayley:martin:96,barth:bayley:98} Heat denaturation
studies have shown that cCaM of Ca$^{2+}$-free (apo) CaM starts to
denature slightly above the physiological
temperature.~\cite{tsalkova:privalov:85} The
denaturation of the apo-cCaM was observed at lower concentration
of denaturant than denaturation of the nCaM, while the order was
reversed for Ca$^{2+}$-CaM.~\cite{masino:bayley:00} From
temperature-jump fluorescence spectroscopy by Rabl et
al.~\cite{rabl:bayley:02} the instability of the cCaM was also
observed from the study of unfolding of apo-CaM. They suggested that
the cCaM was partially unfolded at native conditions.  Recent NMR
experiments done by Lundstrom and Akke~\cite{lundstrom:akke:05} monitoring
relaxation rates involving 
$^{13}$C$^{\alpha}$ spins in adjacent residues of E140Q mutant of cCaM
revealed transient partial unfolding of helix F. This was interpreted as a
global exchange process involves a partially unfolded minor state that 
was not detected previously~\cite{evenas:akke:99, evenas:akke:01}. A very
recent conformational dynamics simulation of the cCaM by Chen et
al.~\cite{gwei:hummer:07} has shown presence of an unfolded apo-state.
These observations further suggests that the conformational exchange may 
be more complex than a simple two-state process.

The variational model presented in this paper agrees with the emphasis
on differences in inherent flexibility of the two domains to
understand their distinct physical characteristics and the complexity
in the conformational transition mechanism.~\cite{chou:bax:01} Our
results also agree with a MD simulation study~\cite{barton:caves:02}
which has shown that the nCaM is inherently more flexible with lower
binding affinity than the cCaM. The more open conformation and lower
intrinsic flexibility of the cCaM is also probably the key to
understanding initial binding between this domain and CaM's target
enzymes.~\cite{barton:caves:02,yamniuk:vogel:04}

The variational model presented in this paper predicts that the different inherent
flexibilities of the two domains of CaM also lead to distinct
transition mechanisms, even though the folded state topology of the
two domains is the same.  The mechanism controlling the open/closed
transition does not involve cracking for the relatively flexible nCaM
whereas the mechanism controlling the conformation transition of the
more rigid cCaM exhibits transient partial local unfolding in its
helix-linker between the binding loops.
This partial local unfolding or cracking in the cCaM shows the 
complexity of the open/closed conformational transition mechanism of cCaM.


A recent NMR experiment studied the EF-hand association affects on the
structure, Ca$^{2+}$ affinity, and cooperativity of
CaM.~\cite{lakowski:mcintosh:07} The EF-hands in CaM domains are
``odd-even" paired with EF-hands 1 and 2 in nCaM and EF-hands 3 and 4
in cCaM. This arrangement of EF-hands in CaM is thought to be a
consequence of its evolution from a biologically related ancestor
EF-hand by gene duplication.~\cite{nakayama:kretsinger:92} The
``even-odd" pairing of EF-hands 2 and 3 (CaM2/3)
(Fig.~\ref{fig:cam_domains}(c)) has been characterized by NMR in
Ref.~\onlinecite{lakowski:mcintosh:07}.  In this fragment, EF-hands 2
and 3 are connected by the central linker between nCaM and
cCaM. Although, from the crystal structure of holo Ca$^{2+}$-CaM this
interdomain linker region is observed to be a long rigid
$\alpha$-helix,~\cite{babu:cook:85} several NMR relaxation experiments
have demonstrated that this central linker is flexible in solution
near its midpoint and the two domains do not
interact.~\cite{barbato:bax:92,spoel:vogel:96}  In contrast to the high affinity
and positive cooperativity for Ca$^{2+}$ binding in the two
``odd-even" paired EF-hand domains of CaM, the CaM2/3 binds Ca$^{2+}$
in a sequence. First in the high-affinity EF-hand 3 and then in the
EF-hand 2 with much lower affinity.~\cite{lakowski:mcintosh:07}
Although not a direct focus of our paper, we also note that a
peptide binding to CaM2/3 has also been characterized recently by 
McIntosh and co-workers.~\cite{lakowski:mcintosh:07a} It was found that 
CaM2/3 adopts Ca$^{2+}$-bound
structure with peptide binding very much similar to those of nCaM or
cCaM. These observations reflect the very high plasticity of the
EF-hand association and mediate Ca$^{2+}$-dependent recognition of
target proteins.

Our results of apo/holo conformational transition of CaM2/3 reveals
that the central linker in CaM2/3 is highly flexible and dominates the
transition mechanism.  The folding of this linker to a rigid
$\alpha$-helix is preceded by a further increase in its flexibility
and local unfolding. This cracking leads to a very sharp transition of
the helix-linker from apo-CaM2/3 to holo-CaM2/3 conformation.  While
our model did not predict partial local unfolding or cracking of
helix-F from the open/closed conformational transition of cCaM, the
apo/holo conformational transition of CaM2/3 reveals cracking in some
region of helix-F as we have discussed already in comparison with the
NMR study~\cite{lundstrom:akke:05}. The predicted change in
conformational flexibility of CaM2/3 reveals the stepwise binding of
Ca$^{2+}$ ions with binding loop III having higher affinity than
binding loop II.  Furthermore, our results suggests that due to very
high inherent flexibility the central linker in CaM2/3 behaves more
like a helix-linker in CaM domains, as concluded from NMR
measurements.~\cite{lakowski:mcintosh:07} On the other hand, in
Ca$^{2+}$-bound (holo) CaM structure~\cite{chattopadhyaya:quiocho:92}
this same central linker adopts a more stable $\alpha$-helix
conformation which requires cracking in the conformational transition from its apo-CaM
structure.

\section*{METHODS} 

\begin{figure}
   \begin{center}
      \includegraphics[width=3in]{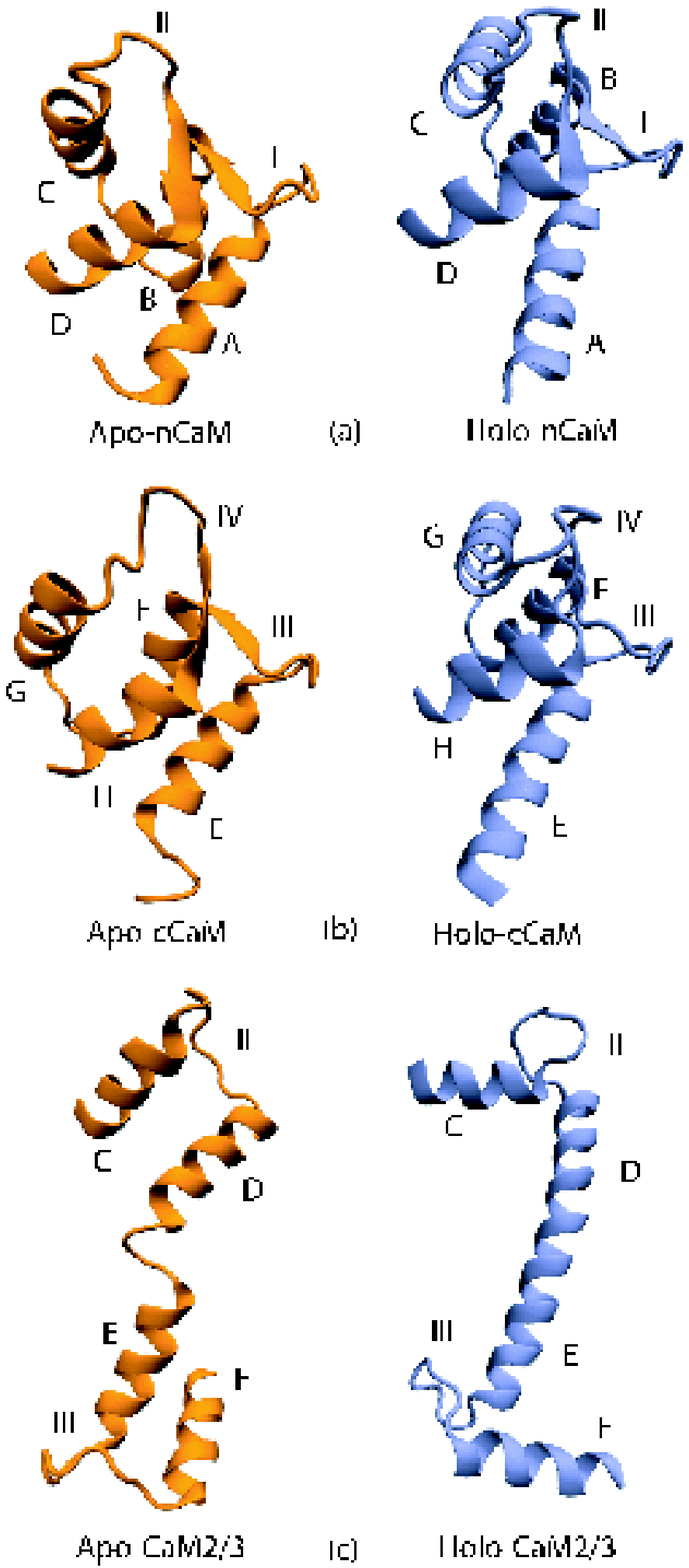}
      \caption{}
      \label{fig:cam_domains}
   \end{center}
\end{figure}

\textbf{Preparation of Structures.}~ The structures of the two
conformations of the transition are rotated to have the same center of
mass and minimum root-mean square deviation of the $C_{\alpha}$
positions.  In nCaM, EF-hands 1 and 2 are consist of helices A/B and
C/D (Fig.~\ref{fig:cam_domains}(a)), respectively, whereas helices E/F
and G/H in cCaM (Fig.~\ref{fig:cam_domains}(b)) form EF-hands 3 and 4,
respectively. The fragment CaM2/3 contains EF-hands 2 and 3
with helices C/D and E/F (Fig.~\ref{fig:cam_domains}(c)),
respectively.  The conformational transition in cCaM is modeled from 
the residues 76-147 of apo-CaM (PDB ID {\em
1cfd})~\cite{kuboniwa:bax:95} and holo-CaM (PDB ID {\em 1cll})
~\cite{chattopadhyaya:quiocho:92} structures, while the conformational
transition in nCaM is modeled from the residues 4--75 of the same PDB 
structures. The apo to holo 
conformational transition of the ``even-odd" paired EF-hand motifs
CaM2/3 are modeled from EF-hands 
2 and 3, residues 46-113 of apo-CaM (PDB ID {\em 1cfd}) and holo-CaM 
(PDB ID {\em 1cll}) structures.

\textbf{The Variational Model of Conformational Transitions.}~ A
conformation of a protein in our model (described in more detail in
Ref.\onlinecite{tripathi:portman:08}) is described by the $N$ position
vectors of the $\alpha$-carbons of the polypepetide backbone,
$\{\mathbf{r}_i\}$. Partially ordered ensembles of polymer
configurations are described by a coarse-grained reference Hamiltonian
\begin{equation} \label{eq:model}
 \mathcal{H}_0/k_{\mathrm{B}}T = \mathcal{H}_{\mathrm{chain}}/k_{\mathrm{B}}T +
  \frac{3}{2a^2} \sum_{i}C_i[\mathbf{r}_i -
  \mathbf{r}_i^{N}(\alpha_i)]^2,
\end{equation}
where $T$ is the temperature and $k_{\mathrm{B}}$ is Boltzmann's
constant.  Here, $\mathcal{H}_{\mathrm{chain}}$ is a harmonic
potential that enforces chain connectivity of a freely rotating
polymer with mean bond length $a = 3.8$\AA,~\cite{portman:wolynes:01a}
and mean valance angle $\cos \theta = 0.8$.\cite{bixon:zwanzig:78} The
second term includes $N$ variational parameters, $\{C_i\}$, that control
the magnitude of the fluctuations about $\alpha$-carbon position
vectors 
\begin{equation}
\mathbf{r}_i^{N}(\alpha_i) = \alpha_i \mathbf{r}_i^{N_{\mathrm{apo}}}
+ (1 - \alpha_i)\mathbf{r}_i^{N_{\mathrm{holo}}}.
\end{equation}
Here, we have introduced another set of $N$ variational parameters, $\{\alpha_i \}$ ( ranging between
0 and 1), that specify the backbone positions as an
interpolation between the apo-CaM and holo-CaM conformations,
$\{\mathbf{r}_i^{N_{\mathrm{apo}}}\}$ and
$\{\mathbf{r}_i^{N_{\mathrm{holo}}}\}$, respectively. The probability
for a particular configurational ensemble specified by the variational
parameters $\{C_i,\alpha_i\}$ at temperature $T$ is given by the
variational free energy $F(\{C\},\{\alpha\}) = E(\{C\},\{\alpha\}) -
TS(\{C\},\{\alpha\})$. Here, $S(\{C\},\{\alpha\})$ is the entropy loss
due to the localization of the residues around the mean positions
$\{\mathbf{r}_i(\alpha_i)\}$.  The energy is determined by the
two-body interactions between distant residues $E(\{C\},\{\alpha\}) =
\sum_{[i,j]} \epsilon_{ij} u_{ij}$, where $u_{ij}$ is the average of
the pair potential $u(r_{ij})$ over
$\mathcal{H}_0$,~\cite{portman:wolynes:01a} and $\epsilon_{ij}$ is the
strength of a fully formed contact between residues $i$ and $j$ given
by Miyazawa-Jernigan interaction
parameters.\cite{miyazawa:jernigan:96} The sum is restricted to a set
of contacts $[i,j]$ determined by pairs of residues in the proximity
in each of the meta-stable conformations.~\cite{tripathi:portman:08}
Each meta-stable conformation has a distinct but overlapping set of
contacts $[i,j]_{\mathrm{apo}}$, $[i,j]_{\mathrm{holo}}$ and $[i,j]$
is the union of these sets of contacts. In this model, the interaction
energy for contacts that occur exclusively in only one meta-stable
structure is given by $\exp(-u_{ij}/k_\mathrm{B}T) = 1 + \exp(-\langle
u_{ij} \rangle_0/k_\mathrm{B}T)$.  This two-state model provides an
interpolation at the individual contact level of the interaction
energy determined by the two meta-stable conformations and is similar
to some other native state biased potentials describing conformational
transitions.\cite{best:hummer:05,maragakis:karplus:05,okazaki:wolynes:06}

Analysis of the free energy surface parameterized by $\{C,\alpha\}$
follows the program developed to describe
folding:\cite{portman:wolynes:01a} the mechanism controlling the
kinetics of the transitions is determined by the ensemble of structures
characterized by the monomer density at the saddlepoints of the free
energy. At this point, we simplify our model and restrict the
interpolation parameter $\alpha_i$ to be the same for all residues,
$\alpha_i = \alpha_0$.\cite{kim:chirikjian:02} With this
simplification, the numerical problem of finding saddle-points with
respect to $\{C,\alpha\}$ simplifies to minimizing the free energy
$F(\{C\},\alpha_0)$ with respect to $\{C\}$ for a fixed $\alpha_0$.

\textbf{Conformational Flexibility.}~ Main-chain conformational
flexibility is characterized by the mean-square fluctuations
$\{\mathcal{B}_i(\alpha_0)\}$ of each $\alpha$-carbon of the
polypepetide chain from its average positions, $\delta \mathbf{r}_i$,
along the transition route as $\alpha$ goes from 0 to
1.~\cite{tripathi:portman:08} These natural order
parameters for the reference Hamiltonian $\mathcal{H}_0$, $\mathcal{B}_i =
\langle \delta \mathbf{r}_i^2\rangle_0$, 
contains information about the degree of structural order of each
residue.~\cite{portman:wolynes:98,portman:wolynes:01a}

\textbf{Conformational Transition Mechanism.}  The main-chain dynamics
responsible for the detailed mechanism of the conformational transition
in our analysis is based on the order parameters introduced to
describe the folding mechanism $\rho_i^{\mathrm{apo/holo}} = \langle
\exp(-\alpha^{\mathrm{N}}(\mathbf{r}_i -
\mathbf{r}_i^{\mathrm{N}_\mathrm{apo/holo}})^2\rangle_0$ with
$\alpha^{\mathrm{N}} = 0.5$ defining the width of a Gaussian window
about the meta-stable structure
$\{\mathbf{r}_i^{\mathrm{N}_\mathrm{apo/holo}}\}$.  In particular, it is
convenient to characterize the relative similarity to the apo
structure along the transition route through the normalized measure
\begin{equation} \label{eq:norm density 1}
  \overline{\rho_i}^{\mathrm{apo}}(\alpha_0) = 
  \frac{\rho_i^{\mathrm{apo}}(\alpha_0) - \rho_i^{\mathrm{apo}}(0)}
       {\rho_i^{\mathrm{apo}}(1) - \rho_i^{\mathrm{apo}}(0)},
\end{equation}
where $\rho_i^{\mathrm{apo}}(\alpha_0)$ is the monomer density of the
$i^{\mathrm{th}}$ residue with respect to the apo conformation
described by
$\{\mathbf{r}_i^{N_{\mathrm{apo}}}\}$~\cite{tripathi:portman:08}.
Similarly, we represent the relative structural similarity to the holo
conformation as
\begin{equation} \label{eq:norm density 2}
  \overline{\rho_i}^{\mathrm{holo}}(\alpha_0) 
  = \frac{\rho_i^{\mathrm{holo}}(\alpha_0) - \rho_i^{\mathrm{holo}}(1)}
           {\rho_i^{\mathrm{holo}}(0) - \rho_i^{\mathrm{holo}}(1)},
\end{equation}
where $\rho_i^{\mathrm{holo}}(\alpha_0)$ is the monomer density of the
$i^{\mathrm{th}}$ residue with respect to the holo conformation
described by $\{\mathbf{r}_i^{N_{\mathrm{holo}}}\}$.  In the holo
state, $\overline{\rho_i}^{\mathrm{apo}}(0)=0$ and
$\overline{\rho_i}^{\mathrm{holo}}(0)=1$, while in the apo state
$\overline{\rho_i}^{\mathrm{apo}}(1)=1$ and
$\overline{\rho_i}^{\mathrm{holo}}(1)=0$.  To represent the structural
changes more clearly, it is convenient to consider the difference,
\begin{equation}\label{eq:natdens_diff}
\Delta\overline{\rho_i}(\alpha_0) =
\overline{\rho_i}^{\mathrm{apo}}(\alpha_0) - \overline{\rho_i}^{\mathrm{holo}}(\alpha_0) 
\end{equation}
for each residue. This difference shifts the relative degree of
localization to be between $\Delta\overline{\rho_i}(1) = 1$ and
$\Delta\overline{\rho_i}(0) =-1$ corresponding to the apo and holo
conformations, respectively.

\newpage

\textbf{REFERENCES}


\clearpage 

\section*{Figure Legends}

\subsubsection*{Figure~\ref{fig:cam_domains}.}
Three dimensional structures of calmodulin (CaM) domains and EF-hands 2 and 
3 fragment. The apo-CaM and holo-CaM structures shown here 
are correspond to human CaM with PDB code {\em 1cfd} and {\em 1cll} respectively.
(a) The closed, apo and open, holo conformations of N-terminal domains of CaM (nCaM) 
consist of helices A/B and C/D with binding loops I and II respectively. 
(b) The closed, apo and open, holo conformations of C-terminal domains of CaM (cCaM) 
consist of helices E/F and G/H with binding loops III and IV respectively. 
(c) The unfolded (apo) and folded (holo) EF-hands of CaM (CaM2/3) consist 
of helices C/D and E/F with binding loops II and III respectively. 
These three-dimensional illustrations were made using Visual Molecular 
Dynamics (VMD).\cite{humphrey:schulten:96}

\subsubsection*{Figure~\ref{fig:B_rho}.}
(a) Fluctuations $\mathcal{B}_i$ vs residue index of N-domain of CaM 
(nCaM) for selected values of the interpolation parameter $\alpha_0$ 
in the conformational transition route between open and closed structures. 
The secondary structure of nCaM is indicated above the plot. Helices are 
represented by the rectangular boxes, binding loops and helix-linker are 
by lines and small $\beta$-sheets are by arrows.
(b) Difference between the normalized native density
$\Delta\overline{\rho_i}$ (a measure of structural similarity) of each
residue for different $\alpha_0$. The change in color from red to blue
is showing the closed to open conformational transition of
nCaM. This is normalized to be $-1$ at the open state minimum
($\alpha_0=0$; blue) and 1 at the closed state minimum ($\alpha_0=1$;
red).

\subsubsection*{Figure~\ref{fig:B_rho_1}.}
(a) Fluctuations $\mathcal{B}_i$ vs residue index of C-domain of CaM 
(cCaM) for selected values of the interpolation parameter $\alpha_0$ 
in the conformational transition route between open and closed structures. 
The secondary structure of cCaM is indicated above the plot. 
(b) Difference between the normalized native density
$\Delta\overline{\rho_i}$ (a measure of structural similarity) of each
residue for different $\alpha_0$. The change in color from red to blue
is showing the closed to open conformational transition of
cCaM. This is normalized to be $-1$ at the open state minimum
($\alpha_0=0$; blue) and 1 at the closed state minimum ($\alpha_0=1$;
red).

\subsubsection*{Figure~\ref{fig:B_domains}.}
Change in fluctuations $\mathcal{B}_i$ of the residues from binding loops and 
helix-linker of CaM domains along the open/closed conformational transition 
route for different $\alpha_0$. Residues Gly23, Gly59 and Glu45 are from binding 
loop I, loop II and the helix-linker of the nCaM, respectively. While, residues 
Gly96, Gly132 and Asp118 are from binding loop III, loop IV and the helix-linker 
of the cCaM, respectively.

\subsubsection*{Figure~\ref{fig:uij_domains}.}
Average pair potentials $u_{ij}$ of the residues from the helix-linker of 
CaM domains along the open/closed conformational transition route for 
different $\alpha_0$. Residue Glu45 and Asp118 are from the nCaM and cCaM, 
respectively.
 
\subsubsection*{Figure~\ref{fig:B_rho_2}.}
(a) Fluctuations $\mathcal{B}_i$ vs residue index of fragment EF-hands 2 
and 3 of CaM (CaM2/3) for selected values of the interpolation parameter 
$\alpha_0$ in the conformational transition route between unfolded and 
folded states. The secondary structure of CaM2/3 is indicated above the 
plot.
(b) Difference between the normalized native density
$\Delta\overline{\rho_i}$ (a measure of structural similarity) of each
residue for different $\alpha_0$. The change in color from red to blue
is showing the unfolded to folded conformational transition of
CaM2/3. This is normalized to be $-1$ at the folded state minimum
($\alpha_0=0$; blue) and 1 at the unfolded state minimum ($\alpha_0=1$;
red).

\begin{figure}
   \begin{center}
      \includegraphics[width=6in]{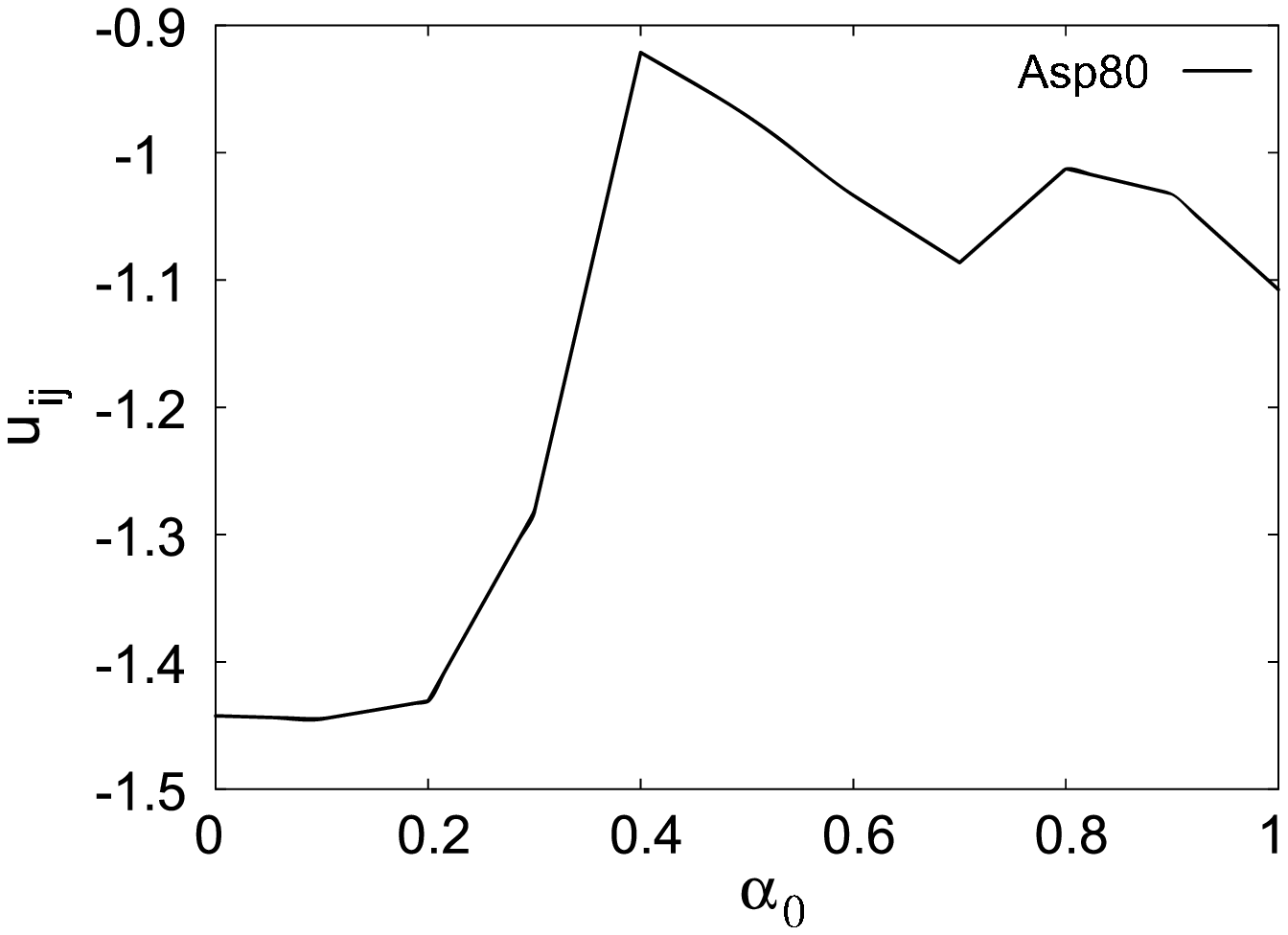}
      \caption{}
      \label{fig:uij_cam2/3}
   \end{center}
\end{figure}

\subsubsection*{Figure~\ref{fig:uij_cam2/3}.}
Average pair potentials $u_{ij}$ of the residues Asp80 from the central linker of 
CaM along the open/closed conformational transition route for different $\alpha_0$. 

\clearpage

\end{document}